# Streamline-Directed Tunable Deterministic Lateral Displacement (DLD) Chip: A Novel Approach to Efficient Particle Separation

Ali Kheirkhah Barzoki[1], Amir Shamloo[1,2,*]

[1]Department of Mechanical Engineering, Sharif University of Technology, Tehran, Iran

[2]Stem Cell and Regenerative Medicine Center, Sharif University of Technology, Tehran, Iran

* Correspondence: shamloo@sharif.edu

## Abstract

In conventional Deterministic Lateral Displacement (DLD), the migration behavior of a particle of specific size is determined by the critical diameter ($D_c$), which is predefined by the device's geometry. In contrast to the typical approach that alters the angle between the pillar array and fluid streamlines by modifying the geometrical parameters, this study introduces a novel perspective that focuses on changing the direction of the streamlines. The proposed technique enables the fabrication of a tunable DLD chip that is both easy to manufacture and design. This chip features one completely horizontal pillar array with two bypass channels on the top and bottom of the DLD chamber. The width of these bypass channels changes linearly from their inlet to their outlet. Two design configurations are suggested for this chip, characterized by either parallel or unparallel slopes of the bypass channels. This chip is capable of generating a wide range of $D_c$ values by manipulating two distinct control parameters. The first control parameter involves adjusting the flow rates in the two bypass channels. The second control parameter entails controlling the slopes of these bypass channels. Both of these parameters influence the direction of particle-carrying streamlines resulting in a change in the path-line of the particles. By changing the angle of streamlines with pillar array, the $D_c$ can be tuned. Prior to determining the $D_c$ for each case, an initial estimation was made using a Python script that utilized the streamline coordinates. Subsequently, through FEM modeling of the particle trajectories, precise $D_c$ values were ascertained and juxtaposed with the estimated values, revealing minimal disparities. This innovative chip enables the attainment of $D_c$ values spanning from 0.5 to 14 μm.

# 1. Introduction

Microfluidics initially emerged to fulfill the need for biomolecular analysis, but in recent years, its scope has expanded to encompass cell separation studies (McGrath et al., 2014). Cell separation and manipulation are pivotal steps in various biological and medical assays, and microfluidic devices offer distinct advantages for working at the cellular level (Choe et al., 2021). With features like low Reynolds numbers, predictable flows, small dimensions, and minimal fluid volumes, along with well-established microfabrication techniques and materials, these devices enable precise control and handling (Sun et al., 2020). Microfluidic separation methods can be classified as active or passive (Song et al., 2023). Active methods involve the application of external forces, while passive methods rely on carefully designed channel geometries and internal forces to sort different types of particles. Several common active separation methods include dielectrophoresis, magnetophoresis, acoustophoresis, immunomagnetic separation (IMS), flow cytometry, and optical force (Ashkezari et al., 2022; Chen et al., 2020; S. Liu et al., 2019; Momeni et al., 2023; Nasiri et al., 2022; Voronin et al., 2020). On the other hand, passive methods employ various techniques, such as incorporating pillars, weirs, and objects within microchannels, pinched-flow fractionation (PFF), hydrodynamic filtration (HDF), inertial forces, and biomimetic separation (Chiu et al., 2016; Dezhkam et al., 2023; Ebrahimi et al., 2023; Keumarsi et al., 2023; Pødenphant et al., 2015). These methods take advantage of specific particle properties, including size, shape, deformability, compressibility, density, as well as dielectric, magnetic, and adhesive properties to achieve effective separation.

A high-throughput passive particle separation method known as deterministic lateral displacement (DLD) has shown great promise for size-based particle separation. The method employs an array of pillars arranged in a specific configuration to separate the flow streamlines, leading to the segregation of particles based on a critical diameter value, $D_c$. This separation relies on the steric interaction between the flowing particles and the pillar arrays. Larger particles, exceeding the $D_c$ value, tend to travel along pillar rows, resulting in a cross-flow motion upon colliding with the pillars. Conversely, smaller particles, below the $D_c$ threshold, follow the streamlines without deviating from their path. Initially reported by Huang et al. in 2004, this technique allows the separation of particles based on their sizes in continuous flow with a remarkable resolution of down to 10 nm (Huang et al., 2004). Since its introduction, DLD has found extensive applications in various fields, accommodating particles ranging from millimeters to micrometers and even submicrometer sizes (Balvin et al., 2009; Beech et al., 2009; Z. Liu et al., 2021; Tottori & Nisisako, 2020; Zheng et al., 2005). DLD has particularly gained prominence in medical-related areas, demonstrating successful separation of DNA, white blood cells (WBC), red blood cells (RBC), platelets, and circulating tumor cells (CTC) from blood samples (Inglis et al., 2008; Z. Liu et al., 2021; Ström et al., 2022; Tang et al., 2022; Zheng et al., 2005). Moreover, numerical studies have investigated DLD separation in moderate to high Reynolds number (Re) regimes. The findings indicate that the $D_c$ and particle behavior change with Re; as the Re increases, the $D_c$ decreases (Dincau et al., 2018). Additionally, research has shown that particle deformation and stiffness play a crucial role in DLD separation (Holmes et al., 2014; Wang et al., 2013). In a numerical study by Khodaee et al., the stress exerted on particles in various flow conditions, as well as particle trajectories, were investigated (Khodaee et al., 2016).

Traditional DLD has a drawback in that each DLD array is effective only within a narrow range of predetermined particle sizes. This limitation arises from the fixed $D_c$ in the device's geometry, which determines the migration of particles through the pillar array. However, when dealing with biological particles, their sizes can vary significantly under different experimental conditions, posing challenges. Additionally, fouling and non-specific adsorption on the channel surface can alter the $D_c$ over time, leading to a degradation in the device's separation performance (Tottori & Nisisako, 2023). As a result, there is a growing demand for methods capable of actively modulating the $D_c$ to enable the use of the same DLD device for multiple applications under optimized conditions.

In recent years, researchers have focused on achieving tunable DLD chips. Unlike conventional DLD chips, which have a fixed $D_c$ determined by the microchannel's geometry, tunable DLD chips offer a range of $D_c$ values, making them versatile for various applications. One approach to achieving tunability in DLD chips is by utilizing non-Newtonian fluids as the suspending liquid, as demonstrated in a numerical study by D'Avino (D'Avino, 2013). Another method involves combining dielectrophoresis (DEP) with DLD, as shown by Chang and Cho (Chang & Cho, 2008). They introduced virtual pillar arrays by replacing physical pillars with electrodes, creating an array of negative-DEP obstacles in an electric field. This innovative approach offers tunability to the DLD chip. Calero et al. took a different approach by integrating planar electrodes into a conventional DLD device, producing a force orthogonal to the fluid flow direction (Calero et al., 2019). By utilizing alternating current (AC) electric fields with different frequencies, they successfully enhanced the sorting of particles. Additionally, Tottori and Nisisako implemented pillars made of a thermo-responsive hydrogel that can shrink and swell upon heating and cooling (Tottori & Nisisako, 2023). This unique feature allows for flexible tuning of the $D_c$.

In this study, we adopt a novel perspective on DLD principles. Instead of altering the orientation of the pillar array, we focus on adjusting the streamlines' direction. Our innovation lies in the introduction of a passive tunable DLD separation chip. By manipulating two control parameters - the flow rates of bypass channels and the slope of these channels - we achieve a substantial range of $D_c$ values. This velocity-based tunable chip offers simplicity in controlling the $D_c$ and fabrication, while providing a significant range of $D_c$ values.

## 2. Problem Formulation

DLD is a technique that sorts particles based on their sizes. It involves using micro-posts embedded in a chip to manipulate the flow of particles. When particles flow around these posts, they experience hydrodynamic forces that cause them to move laterally. The key factor is the $D_c$, which determines whether a particle follows the usual flow path or is influenced by the posts. In conventional DLD chips, the posts within a column undergo slight shifts compared to the previous column. This small shift, represented as Δλ, combined with the gap (G) between adjacent posts and the center-to-center distance (λ), collectively ascertain the $D_c$ of separation (see **Fig. 1**). As particles of varying sizes move through the pillar array, they exhibit predetermined pathways. Particles exceeding the $D_c$ are unable to align with the regular flow paths due to their interaction with the pillars. Consequently, they periodically alter their path, transitioning to the "bumped" mode. In this mode, particles move along a path that follows the arrangement of the pillar rows. On the contrary, particles with diameters smaller than the $D_c$ smoothly trace the flow's streamlines, untouched by the presence of the pillars. These smaller particles adopt the "zigzag" pattern, periodically shifting between rows. Additionally, there is another factor to consider – denoted as θ (see **Fig. 1**). This parameter emerges due to the shifting of pillar columns relative to the previous column and indicates the difference between the direction of the pillar array and the streamlines of the fluid flow. Its definition is provided below:

$$\varepsilon = \tan\theta = \frac{\Delta\lambda}{\lambda} \tag{1}$$

For the determination of $D_c$, we employed the empirical equation introduced by Davis (Davis, 2008). This approximation was obtained using 20 different devices, each characterized by varied geometrical parameters and spherical particle sizes, all within the context of a parabolic flow profile. The formula can be expressed as follows:

$$D_c = 1.4G\varepsilon^{0.48} \tag{2}$$

Here, G represents the gap size between neighboring pillars (**Fig. 1**). It should be noted that the geometric values employed in this investigation are 45 μm for G, 90 μm for λ, and 50 μm for $D_p$.

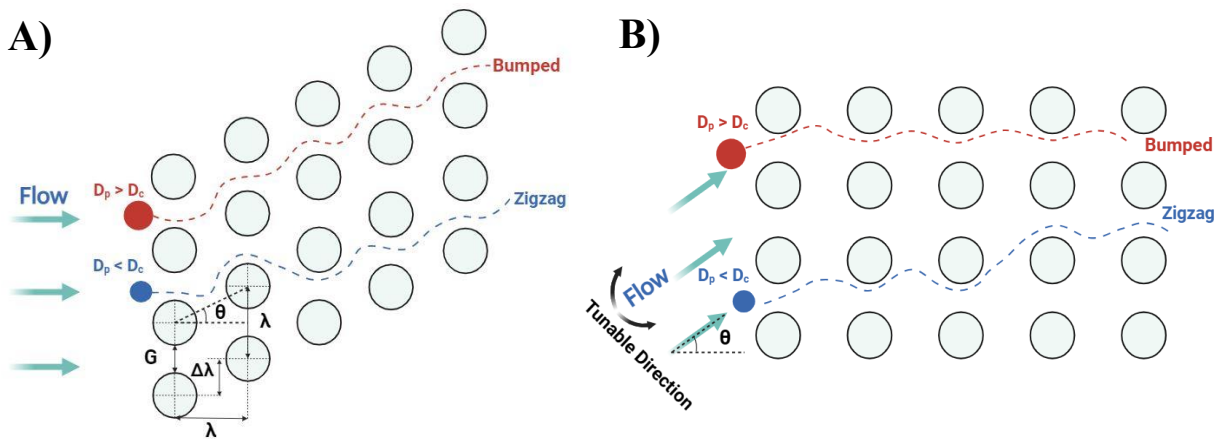

**Figure 1. Schematic illustration of the Geometrical parameters of DLD pillar array. A)** Conventional DLD and **B)** velocity-based tunable DLD concept. λ denotes the center-to-center

distance between the adjacent pillars, $\Delta\lambda$ shows the shift of one column of pillars relative to the previous one, G represents the gap size between two neighboring pillars, $\theta$ is the angle that rows of pillars make with a horizontal line or flow direction, and $D_p$ is the diameter of particles. To be mentioned that the blue particle ($D_p < D_c$) follows the zigzag mode, while the red one ($D_p > D_c$) follows bumped (displacement) mode.

## 2.1  DLD Chip Tunability

Recent studies have explored the potential of making DLD separation chips tunable. Nonetheless, offering a straightforward method for integrating tunability into DLD chips holds promise if it is easily executable. In conventional DLD separation chips, the chip's geometry, especially the angle θ formed between the pillar row and the fluid's streamlines, determines particle separation and sets the $D_c$. Consequently, altering the $D_c$ requires adjustments to these geometric parameters. This implies that a fabricated chip will yield a single specific $D_c$. Existing attempts at creating tunable DLD chips have largely focused on modifying these geometric parameters actively to achieve various $D_c$ values on a single chip. However, viewing DLD principles from a different angle reveals an alternative approach to changing the $D_c$. Instead of altering the direction of the pillar rows (θ) and the geometrical parameters, the direction of the streamlines can be adjusted (see **Fig. 1**). This new perspective allows for flexible alteration of the angle between the pillar rows and streamlines, while all geometric parameters remain constant. By implementing this concept, a simple tunable DLD separation chip can be obtained. A simple chip with a horizontal array of pillars (θ equals 0) in which by regulating the two control parameters, various $D_c$ values can be obtained.

As evident in **Fig. 2**, there are two control parameters by which the $D_c$ can be adjusted. The first parameter is the slope of bypass channels. This chip has two bypass (lateral) channels on top and bottom of the main DLD chamber. Each of these bypass channels has a different inlet and outlet width. This difference forms the slope of the bypass channels. By changing the width of the inlet and outlet of these channels and consequently their slope, the direction of the streamlines coming out of the main inlet. The change in the streamlines' direction results in a change in the path-line of the particles, and consequently the $D_c$ value can be adjusted. It should be noted that this control parameter is a geometrical parameter. Therefore, it cannot be adjusted after the fabrication of the chip. However, it can be set before fabrication based on the range of the $D_c$ that is needed in a specific application. Then, by adjusting the second control parameter, flow rate of bypass channels, the $D_c$ value can be adjusted. To be more specific, this control parameter is to adjust the $D_c$ range before the fabrication, not the exact value of $D_c$.

In this study we have two configurations for the bypass channels. In both configurations, the amount of the slope of the bypass channels is the same. However, the difference of these configurations is the direction of the slope of the top and bottom bypass channels. In the first configuration, both bypass channels have the same slope with the same direction (**Fig. 2A**), while in the second configuration, the bypass channels have the same slope with different directions (**Fig. 2B**). To achieve different slopes, the width of the larger side (inlet or outlet) of the bypass channels gets the constant value of $100 \mu m$ while the smaller side, gets the values of 20, 40, 60, and $80 \mu m$. For each case, we will indicate the slope of the bypass channels, like 100-40 μm.

The second control parameter is the flow rates of the bypass channels. After setting the proper slope for the bypass channels, by adjusting the flow rates of these channels, the direction of streamlines and consequently the $D_c$ can be tuned. In this study, in all cases, the flow rates of the main inlet through which particles come to the DLD chamber and bottom bypass channel were set to 1 $mm/s$ and 10 $mm/s$, respectively. For the top bypass channel, the flow rate got the values of 4, 6, and 8 $mm/s$. The difference between the flow rates of the top and bottom bypass channels can adjust the streamlines direction and the $D_c$. For each case, we will indicate the flow rate ratio of the bypass channels, like 10-4.

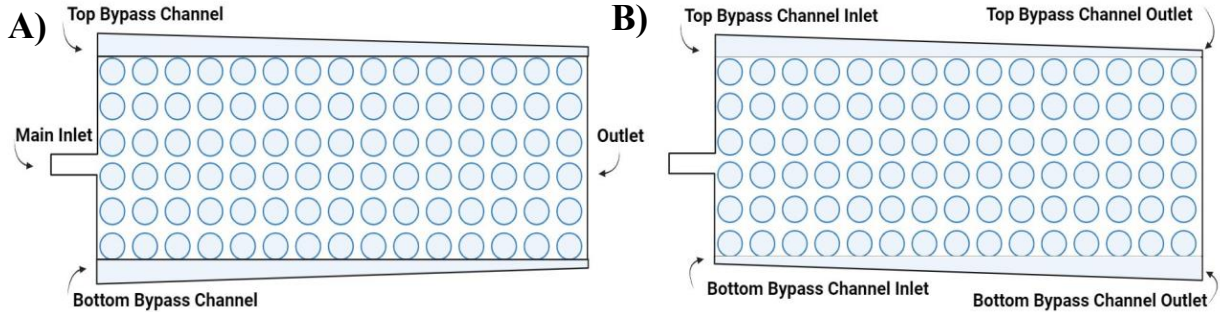

**Figure 2. Schematic illustration of the Geometrical structure of the tunable DLD chip.** The two configurations of the bypass channels, **A)** unparallel-slope configuration and **B)** parallel-slope configuration, are demonstrated. To be noted that there is no wall between the bypass channels and the DLD chamber.

## 2.2    Computational Model and Governing Equations

The trajectory of particles within the fluid was tracked using a particle tracing model based on the Finite Element Method (FEM) using 2D geometries. The fluid equations were solved using the Multifrontal Massively Parallel Sparse direct (MUMPS) Solver, with a set relative tolerance of 1E-4. After solving the fluid and pressure domain, particle trajectories were derived by solving Eq. (5). To determine particle trajectories, the Generalized Minimal Residual Method (GMRES) with left preconditioning was employed, accompanied by residual and relative tolerances of 1E-3 and 1E-6, respectively. Pressure discretization utilized first-order elements, while velocity discretization involved second-order elements. Temporal discretization was achieved through the application of the generalized alpha method. The Newton method was utilized to linearize the set of non-linear equations within each time step. The discretization of the entire domain used triangular elements, excluding the boundary layer elements, which were quadrilateral. The model was developed to account for both fluid inertia and viscosity, which play pivotal roles when dealing with finite Reynolds numbers. To address fluid dynamics, we solved the continuity equation in conjunction with the Navier-Stokes equation, as expressed below:

$$\nabla . \mathbf{v} = 0 \tag{3}$$

$$\rho_f (\mathbf{v} . \nabla)\mathbf{v} = \nabla . [-p\mathbf{I} + \mu(\nabla\mathbf{v} + (\nabla\mathbf{v})^T)] \tag{4}$$

The bold symbols in the equations indicate vectors, while $\rho_f$, $\mathbf{v}$, p, and μ stand for fluid density, velocity, pressure, and dynamic viscosity, respectively. The identity matrix is designated as $\mathbf{I}$. In this study, the assumption is made that the microfluidic chamber is initially filled with fluid before particles are injected into the chamber. Consequently, the initial fluid comprises a solitary phase, and the particle volumetric fraction is insignificantly low to substantially influence fluid flow. To estimate the path of a particle, it becomes imperative to consider the hydrodynamic forces exerted by the fluid on the particle comprehensively. The particle's trajectory can be anticipated by integrating the force balance on the particle as outlined in the following manner (Maxey & Riley, 1983):

$$m_p \frac{d\mathbf{v}_p}{dt} = \mathbf{F}_{Drag} + \mathbf{F}_{Saffman} \tag{5}$$

where bold symbols denote vectors, and $\mathbf{v}_p$ and $m_p$ represent particle velocity and mass, respectively. The first term on the right side signifies the drag force, which acts in the counter direction to the particle's movement. Given that the particle's slip velocity, describing its relative motion with the fluid, remains relatively minor, and its finite Reynolds number is also low, considering adjustments to the drag coefficient is unnecessary in this context (Goldman et al., 1967). However, as a particle comes close to a wall, the drag force increases, necessitating corrections to accommodate particle-wall interaction (Lin et al., 2000). These corrections can notably influence particle behavior, particularly within DLD chips where each particle's behavior is defined in proximity to pillars. In this investigation, the drag force is divided into two components for motion parallel and perpendicular to the wall, with correction factors $\lambda_\perp$ and $\lambda_\parallel$ applied. The resultant drag components are:

$$\mathbf{F}_{Drag,\parallel} = -6\pi\mu r_p (\mathbf{v} - \mathbf{v}_p) \, \lambda_\parallel \tag{6a}$$

$$\mathbf{F}_{Drag,\perp} = -6\pi\mu r_p (\mathbf{v} - \mathbf{v}_p) \, \lambda_\perp \tag{6b}$$

where $r_p$ stands for particle radius, μ denotes fluid viscosity, $\mathbf{v}$ represents the fluid velocity vector, and $\mathbf{v}_p$ indicates the particle velocity vector. The formulas for $\lambda_\perp$ and $\lambda_\parallel$ are typically derived via the "method of reflections" and are mathematically expressed as power series (Lobry & Ostrowsky, 1996). To be noted that closed analytical forms for the effective wall drag force are generally not available, except in exceptional cases such as a sphere's motion in proximity to a flat wall. Assuming the suspended particle is spherical and considerably smaller than obstacles within the fluid, it is reasonable to apply correction coefficients developed by Happel et al. for a particle near a flat wall (Happel & Brenner, 1983). These coefficients are presented as:

$$\lambda_{\parallel}{}^{-1} \cong 1 - \frac{9}{16}\left(\frac{r_p}{z}\right) + \frac{1}{8}\left(\frac{r_p}{z}\right)^3 - \frac{45}{256}\left(\frac{r_p}{z}\right)^4 - \frac{1}{16}\left(\frac{r_p}{z}\right)^5 \tag{7a}$$

$$\lambda_{\perp}{}^{-1} \cong 1 - \frac{9}{8}\left(\frac{r_p}{z}\right) + \frac{1}{2}\left(\frac{r_p}{z}\right)^3 \tag{7b}$$

where z signifies the distance between the center of the spherical particle and the wall. The second term on the right side of Eq. (5) shows the Saffman force, responsible for the lift force brought about by shear and the wall lift effect stemming from the geometry of DLD (Saffman, 1965). This force can be calculated according to the subsequent formulation (Durst & Raszillier, 1989):

$$\boldsymbol{F}_{Saffman} = (K\nu^{\frac{1}{2}}\rho_f d_{ij})/(\rho_p r_p (d_{lk}d_{kl})^{\frac{1}{4}}) \times (\mathbf{v} - \mathbf{v}_p) \tag{8}$$

where $d_{ij}$ is the deformation tensor, $\rho_p$ stands for particle density, K is a constant (2.954), and $\upsilon$ represents the kinematic viscosity of the fluid. To be noted that the Saffman lift force equation presented in Eq. (8) is solely applicable to particles characterized by small Re numbers. Furthermore, the particle's Re, calculated based on its velocity variance from the fluid, must be lesser than the square root of its Re determined by the shear field—a condition met in this study.

This study employs water as the suspending fluid, characterized by a density of 1000 kg/m³ and a dynamic viscosity of 0.001 Pa.s. Furthermore, the particle density is set at 1100 kg/m³, aligning closely with the density of cells. The dimensions of the DLD chamber are 9400 μm in length and 950 μm in width. It is important to highlight that the maximum Reynolds number (Re) observed across all cases in this study is calculated to be 3.54 in the DLD chamber and 50 in the bypass channels. This value indicates the dominance of laminar flow, thus upholding the validity of the laminar flow assumption. The characteristic length was considered as the gap size (45$\mu m$) in the DLD chamber. For the bypass channels, Reynolds number was calculated for both inlet and outlet of the channels and the maximum value is reported.

## 2.3 Model Validation

To ensure the precision of the simulations, the model was validated by the experimtal results by Li et al.. **Fig. 3a** presents a comparison of particle trajectories. **Fig. 3b** provides a quantified representation of particle trajectories along the channel. To achieve this, we extracted the x and y coordinates of particle trajectories from both our simulation and the experiment, plotting them in

**Fig. 3b**. The result reveals a satisfactory agreement between the numerical and experimental results.

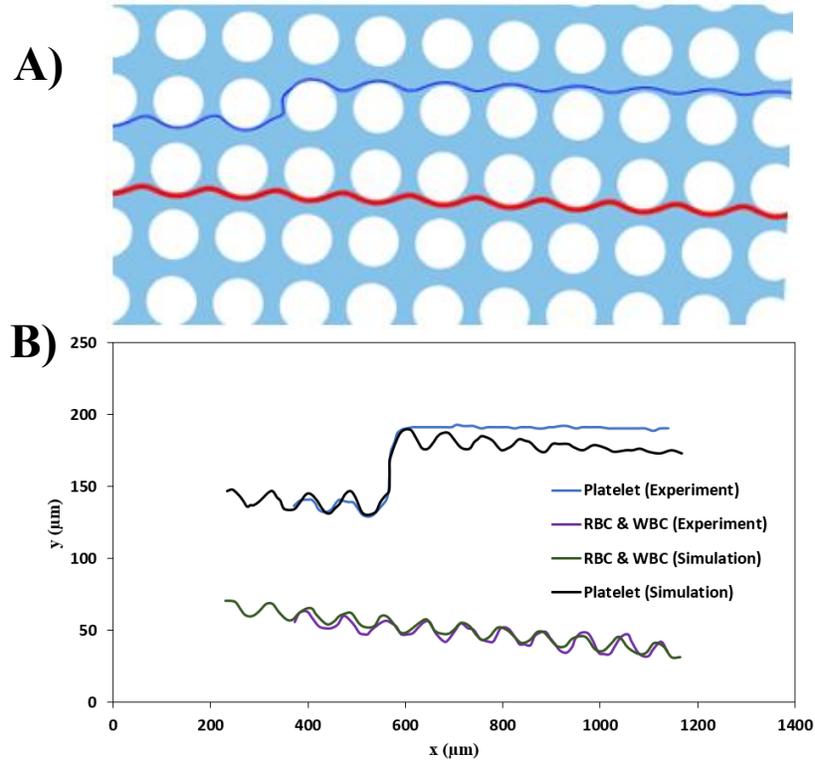

**Figure 3. Model Validation. A)** Separation of platelets (blue) from RBCs and WBCs (red) in our validation simulation. **B)** Quantitative illustration of validation. The x and y coordinates of the particle trajectories were extracted and plotted for both our simulation and the experiment.

## 3. Results and Discussion

Two distinct methods have been explored to ascertain the $D_c$ for each case with varying control parameters. The initial approach involves predicting the $D_c$ through the utilization of streamlines' coordinates extracted from the fluid flow solution, employing Eq. (1) and (2). This method let us calculate the $D_c$ without performing time-consuming FEM simulations to obtain particle trajectories. To validate the $D_c$ prediction in the first approach, for each situation characterized by different control parameters, particle trajectories were obtained via FEM simulation, from which the $D_c$ was then determined. Eventually, results obtained from these two approaches were compared.

### 3.1 $D_c$ Calculation using Python Script

To provide further detail regarding the initial approach, a Python 3.8 script was developed to precisely compute the $D_c$ based on the angle formed by the streamlines and the pillar row (see **Fig. 4**). This involved feeding the extracted streamline coordinates from the fluid flow solution into the Python code. For each individual streamline, the coordinates of the start and end points situated between two adjacent rows of pillars were obtained. Subsequently, employing a linear approximation of the streamline, the slope of the line and the angle at which the streamline intersects the pillar array were determined. This angle θ, coupled with Eq. (1) and (2), facilitated the calculation of the $D_c$ for each pillar row.

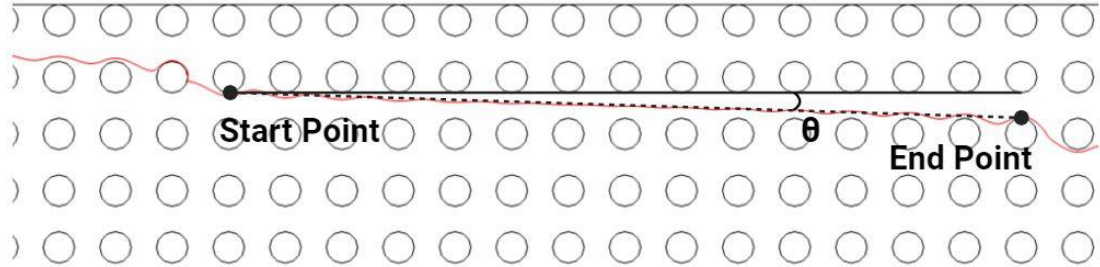

**Figure 4. Visualization of the angle θ that the Python code employs to compute the $D_c$.** The Python code utilizes the streamline's coordinates (red line), fitting a line (dashed line) connecting the start and end points of the streamline within a given row. This process determines the angle θ. This angle mirrors the role of angle θ found in conventional DLD chips.

The effect of each control parameter on the direction of streamlines is depicted in **Fig. 5**. Ten streamlines originating from the main inlet are illustrated. **Figs. 5A-D** show the variation of the direction of streamlines with changing bypass channels slope in the parallel-slope configuration with flow rate ratio of 10-4. A decrease in the slope of the bypass channel causes a reduction in the angle between the streamlines and the pillar array, resulting in a decrease in the value of $D_c$. This trend is consistent across all cases with varying flow rate ratios in both the parallel-slope and unparallel-slope configurations. Thus, setting a specific slope for the geometry enables the tuning of the $D_c$ range prior to fabrication. **Figs. 5E-G** illustrate how the direction of streamlines changes as the flow rate ratios are modified in the parallel-slope configuration, with slope of 100-40 $\mu m$. An increase in the difference between the flow rates of the bypass channels leads to an elevation in the angle between the streamlines and the pillar array. Consequently, the value of $D_c$ also increases.

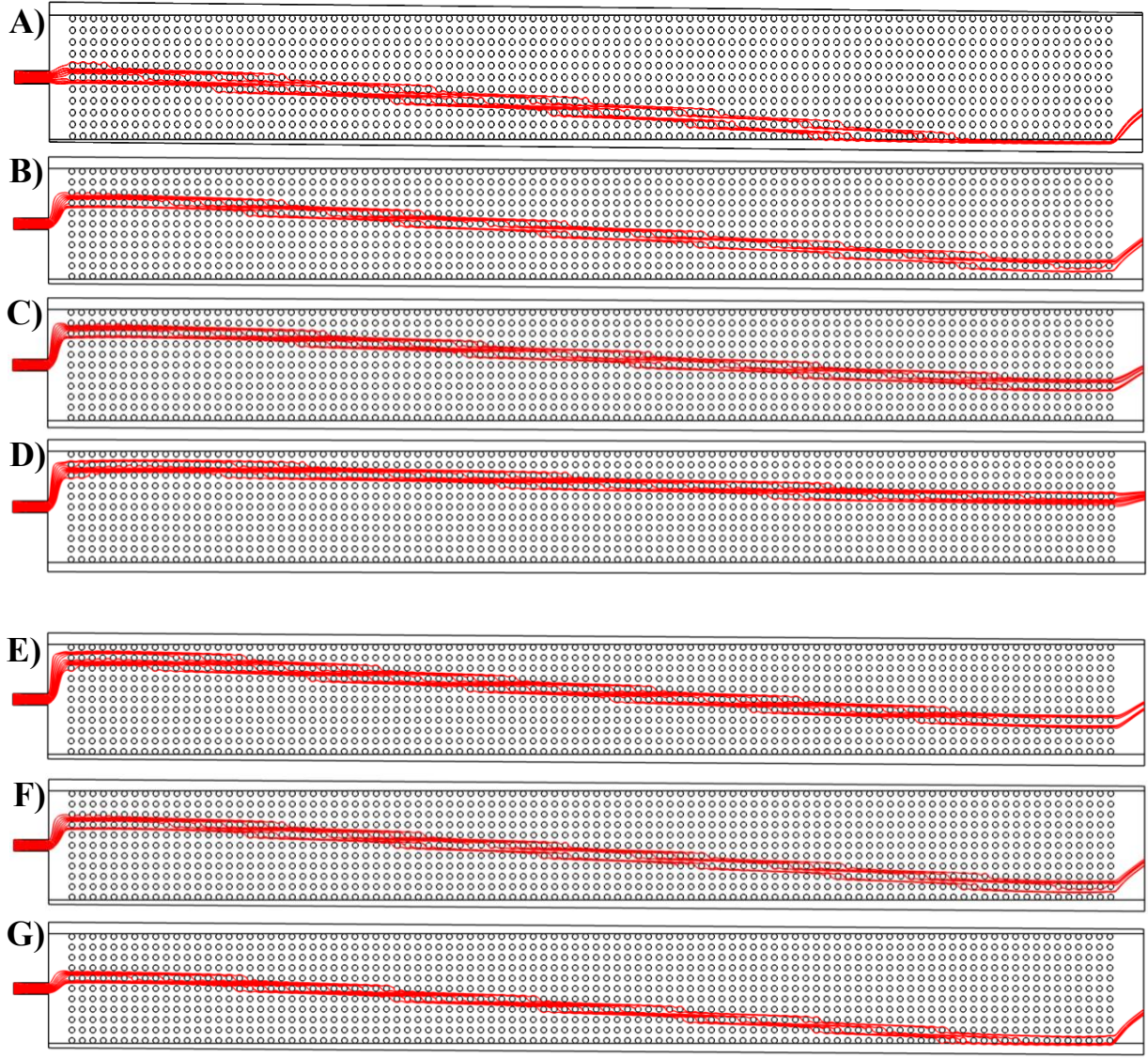

**Figure 5. Changing of the streamline direction with tuning the control parameters in the parallel-slope configuration.** Ten streamlines coming out of the main inlet from different positions of the inlet are depicted (red lines). The direction of the streamlines associated with the bypass slopes of **A)** 100-20 $\mu m$, **B)** 100-40 $\mu m$, **C)** 100-60, and **D)** 100-80 $\mu m$ with flow rate ratio of 10-6 in bypass channels. The direction of the streamlines associated with flow rate ratios of **E)** 10-4, **F)** 10-6, and **G)** 10-8 with bypass channel slopes set to 100-40 $\mu m$.

The variation of the $D_c$ with respect to the two control parameters are depicted in **Fig. 6**. For each case, the calculation of $D_c$ was performed for ten streamlines coming out of the main inlet (see **Fig. 5**). Ten streamlines were considered in order to take into consideration the variations of $D_c$ for all particles coming into the DLD chamber from different position of the main inlet. The error bars demonstrate the standard deviation of $D_c$ for these ten streamlines. **Figs. 6A-C** show the variation of $D_c$ with changing the flow rate ratio and the slope of bypass channels for the unparallel-slope configuration. The general trend in all cases shows that by increasing the flow rate ratio and increasing the slope, the $D_c$ rises. **Figs. 6D-F** demonstrate the $D_c$ variations associated with changing the two control parameters in the parallel-slope configuration. The trend of variations is similar to that of the unparallel-slope configuration. However, the number of bars is more in the parallel-slope configuration. This number corresponds to the number of rows that the streamlines emerging from the main inlet pass through. Therefore, in the parallel-slope configuration, particles change more rows. In other words, they traverse a larger distance across the DLD chamber. As a results, there would be a larger distance between separated particles at the outlet.

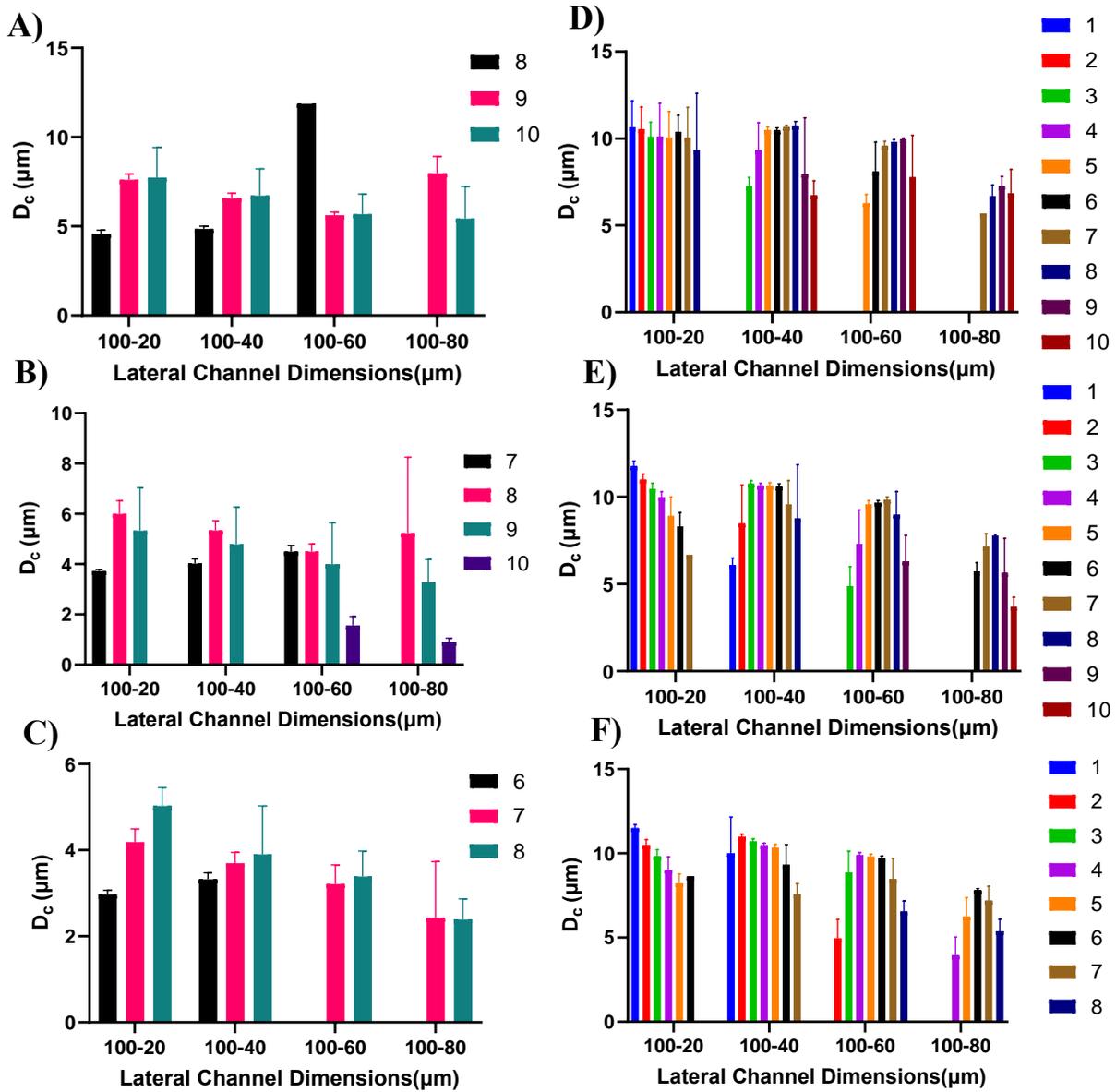

**Figure 6. $D_c$ variations with tuning the control parameters.** The variations of $D_c$ in different bypass channel slopes in the unparallel-slope configuration associated with the bypass channel flow rate ratios of **A)** 10-4, **B)** 10-6, and **C)** 10-8. The variations of $D_c$ in different bypass channel slopes in the parallel-slope configuration associated with the bypass channel flow rate ratios of **D)** 10-4, **E)** 10-6, and **F)** 10-8. Legends show the row number in the DLD chamber, starting from the bottom.

## 3.2    $D_c$ Estimation using FEM Modeling

To evaluate the validation of $D_c$ values calculated by the Python script using streamline coordinates, FEM modeling was performed to find the exact particle trajectories and $D_c$ values. To determine the $D_c$ in each simulation, six distinct particle sizes within a range around the estimated $D_c$ with Python, with a $1 \mu m$ increment, were injected to the DLD chamber. By employing this approach, the precise $D_c$ values were derived through the FEM simulations with a precision of $1 \mu m$.

**Fig. 7** shows the variation of particle trajectories with flow rate ratio in bypass channels with a constant slope for both configurations. It can be inferred that by decreasing the flow rate ratio of the bypass channels in a constant bypass channel slope, the value of $D_c$ decreases. This trend is in agreement with the results obtained from the estimation by Python code (see **Fig. 6**). **Fig. 8** demonstrates the variations of particle trajectories with tuning the slope of bypass channels in a constant flow rate ratio for both configurations. As the slope of bypass channels decreases, the $D_c$ decreases. This is because by lowering the bypass channel angle, the angle of streamlines with the pillar array decreases leading to a fall in the amount of $D_c$. To be noted that here the particles trajectories of just a few cases are demonstrated. However, the results of all cases are illustrated in **Fig. 9**. It is worth noting that by changing the two control parameters, $D_c$ values from about 0.5 to 14 $\mu m$ can be obtained.

An inherent advantage of this chip lies in its step-by-step separation capability. Unlike conventional DLD separation chips where the $D_c$ remains constant across all rows and along the channel, our chip demonstrates varying $D_c$ values both within different rows and along the channel (see **Figs. 7 and 8**). This design permits the sequential separation of particles with differing sizes across multiple stages, channeling them into distinct outlets. As we traverse the channel's length, particles smaller than the $D_c$ gradually transition from the bumped mode to the zigzag mode. This implies that particles of smaller sizes change their rows earlier along the channel (for example see **Fig. 7F**). As a result, not only is separation achieved based on the $D_c$, but particles smaller than the $D_c$ can also be segregated according to their respective sizes along the channel. Furthermore, as particles shift rows, their experience of a different $D_c$ causes another level of size-dependent separation along the channel.

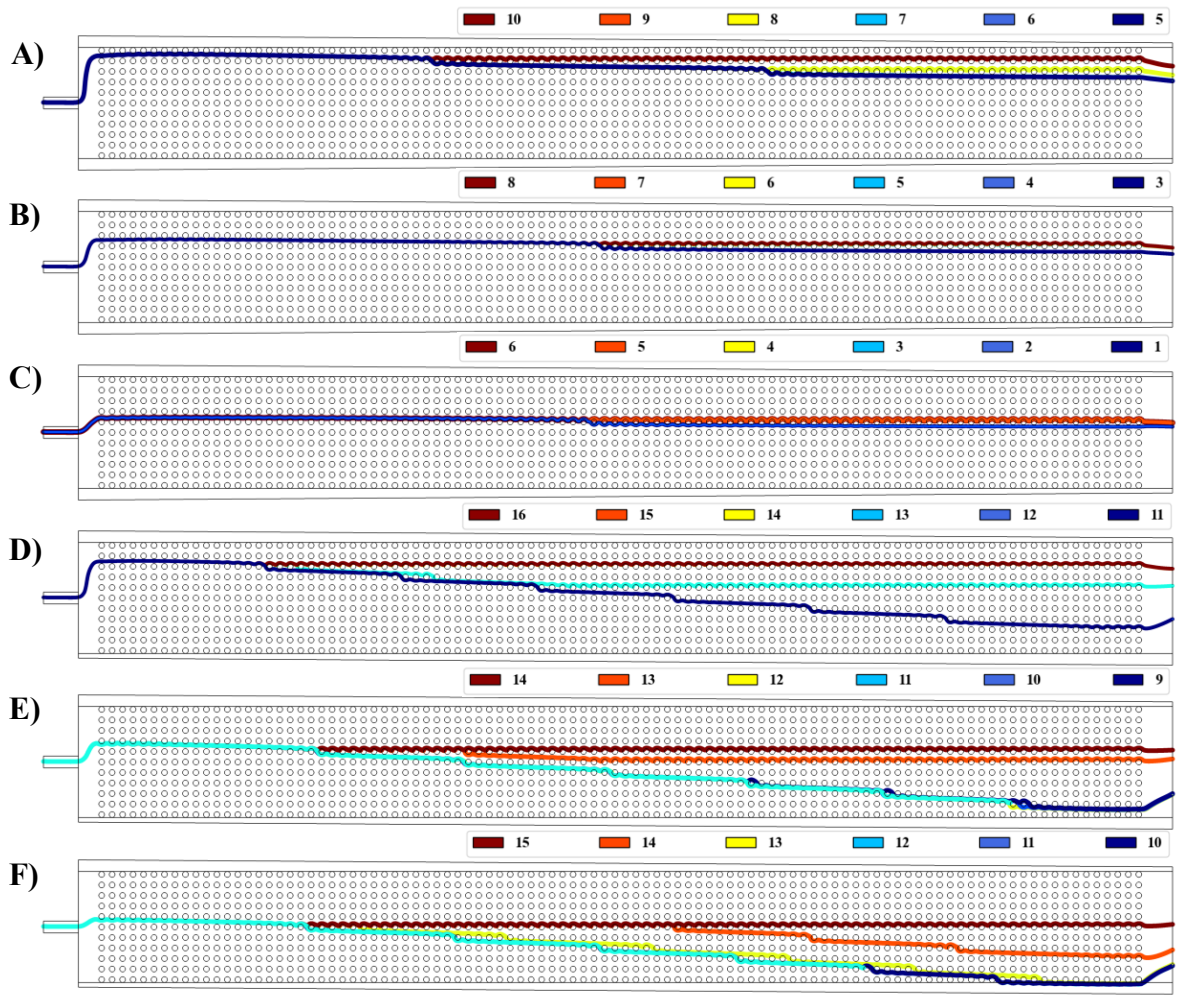

**Figure 7. Variation of particle trajectories with respect to different flow rate ratios in bypass channels.** The separation of particles in the unparallel-slope configuration with flow rate ratios of **A)** 10-4, **B)** 10-6, and **C)** 10-8, and for the parallel-slope configuration with flow rate ratios of **D)** 10-4, **E)** 10-6, and **F)** 10-8 in the bypass channels. The $D_c$ in cases **A-F** is 8, 4, 2, 13, 12, 12 $\mu m$, respectively. In all cases, the slope of bypass channels is 100-40 $\mu m$. Legends show the particle diameter in $\mu m$.

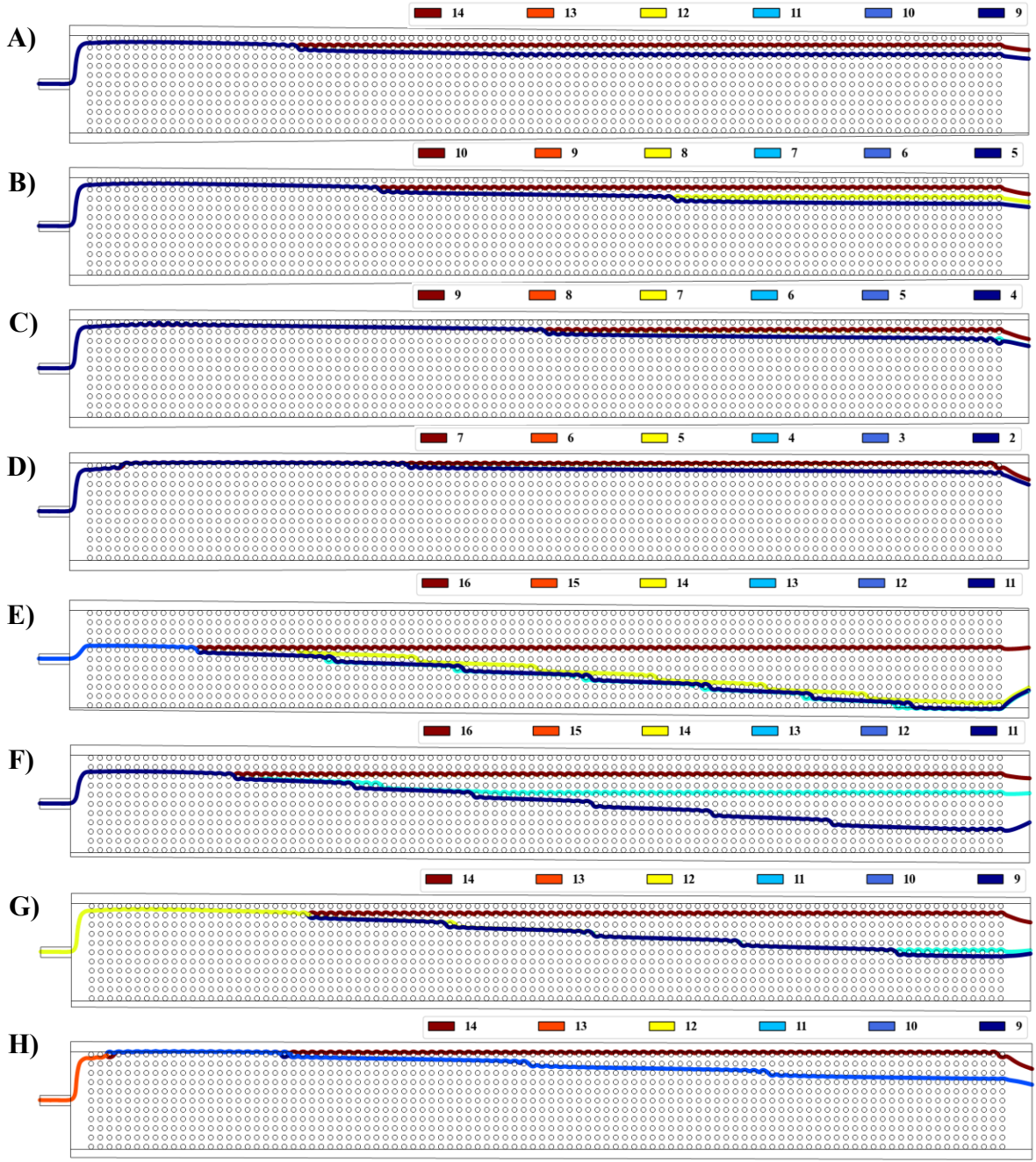

**Figure 8. Variation of particle trajectories with respect to different bypass channel slopes.** The separation of particles for the unparallel-slope configuration with slopes of **A)** 100-20 $\mu m$, **B)** 100-40 $\mu m$, **C)** 100-60 $\mu m$, and **D)** 100-80 $\mu m$, and for the parallel-slope configuration with bypass channel slopes of **E)** 100-20 $\mu m$, **F)** 100-40 $\mu m$, **G)** 100-60 $\mu m$, and **H)** 100-80 $\mu m$. The $D_c$ in cases **A-H** is 10, 8, 6, 2, 13, 13, 12, 10 $\mu m$, respectively. In all cases, the flow rate ratio of bypass channels is 10-4. Legends show the particle diameter in $\mu m$.

## 3.3 Effect of the Channel Length on $D_c$

To examine how the length of the channel impacts the separation process and resulting $D_c$, we analyzed five channels ranging in length from 6000 to 10000 μm. **Fig. 9** displays the changes in $D_c$ as channel length increases. Interestingly, altering channel length does not affect the amount of $D_c$. Additionally, the x position of particle separation remains consistent at approximately 4500 μm for all channel lengths (**Fig. 9A-D**). **Fig. 9E** reveals that the Python code predicts a $D_c$ of approximately 6 μm for all channels. It is important to note that particles with a size equal to or less than 6 μm were separated and changed their row in all cases.

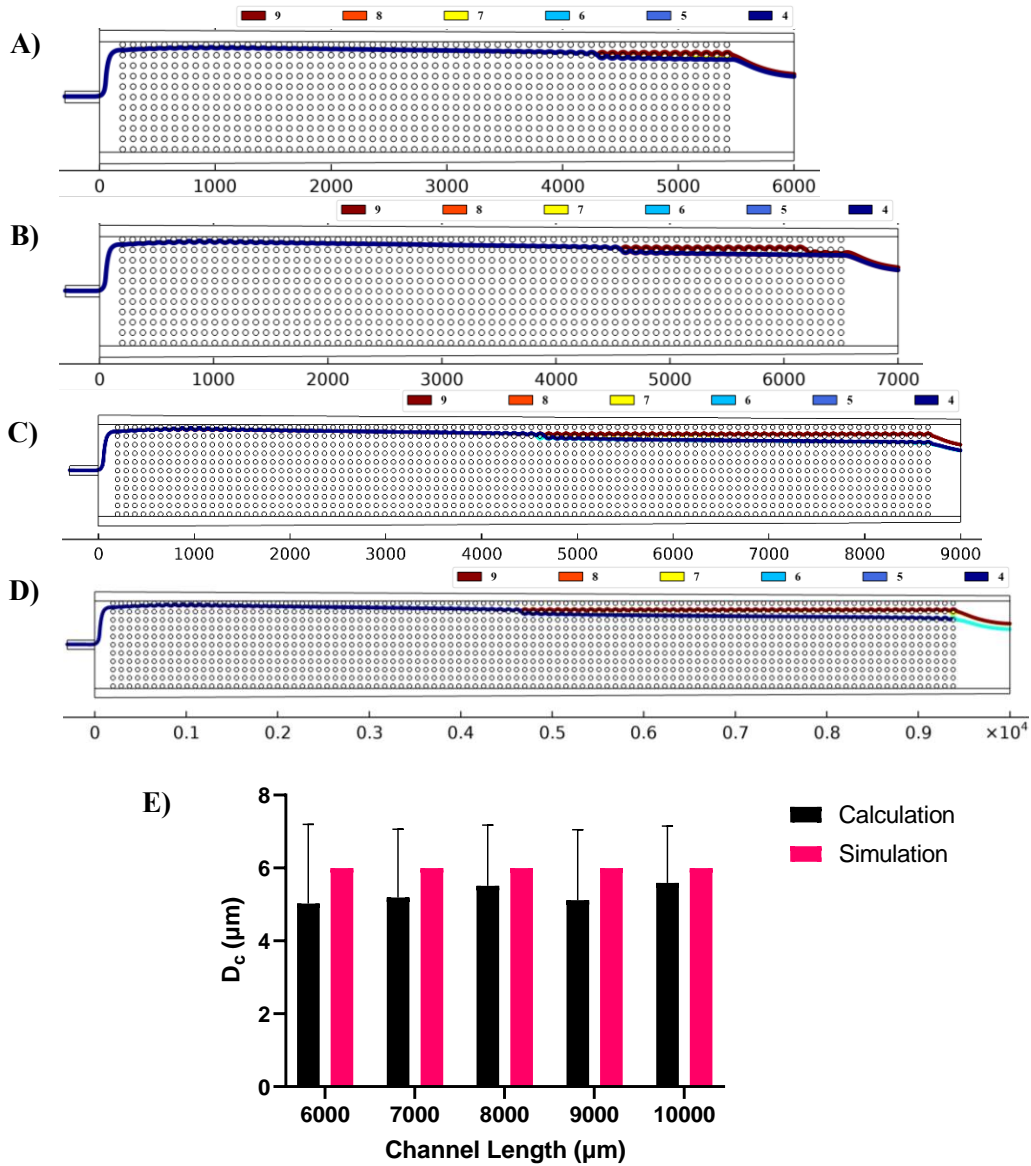

**Figure 9**. $D_c$ **Variations with the channel length.** The separated particles and the separation location in channels with lengths of **A)** 6000 μm, **B)** 7000 μm, **C)** 9000 μm, and **D)** 10000 μm. **E)** Predicted $D_c$ values by Python code and the obtained $D_c$ values by simulation. The error bars show

the standard deviation of $D_c$ for ten streamlines coming out of the main inlet. It should be mentioned that the velocity in the upper and lower bypass channels are 4 and 10 mm/s, respectively. The bypass channel slopes are 100-60 $\mu m$ with unparallel configuration. The color legends illustrate the particle diameter in $\mu m$.

## 3.4    Comparative Analysis of Results from the Two Approaches

**Fig. 10** presents a comparison of the predicted $D_c$ values calculated using the Python code and those obtained from FEM simulations across various control parameters. Clearly, $D_c$ variations have the same trend in both approaches. As the flow rate difference and the slope of the bypass channels increase, the value of $D_c$ increases. Furthermore, the comparison of results for both unparallel-slope (**Figs 10A-D**) and parallel-slope (**Figs 10E-H**) configurations demonstrates that the $D_c$ estimation using the Python script and coordinates of the streamlines has a negligible error with the exact $D_c$ values extracted by the FEM modeling. Therefore, in this method, $D_c$ can be predicted just by having the streamlines obtained from fluid flow simulation without need to perform time-consuming FEM modellings to finds exact particles trajectories.

While a negligible discrepancy exists between the computed $D_c$ and the value derived from the FEM simulation, both exhibit the same trend. This discrepancy arises due to the localized definition of the angle θ between streamlines and the pillar array in each row in our study. In the Python code, this angle is computed as $\arctan(v_y/v_x)$, which $v_y$ and $v_x$ are the vertical and horizontal components of the fluid velocity, respectively. Conversely, in conventional DLD chips governed by Eq. (1) and (2), θ serves as a global geometrical parameter.

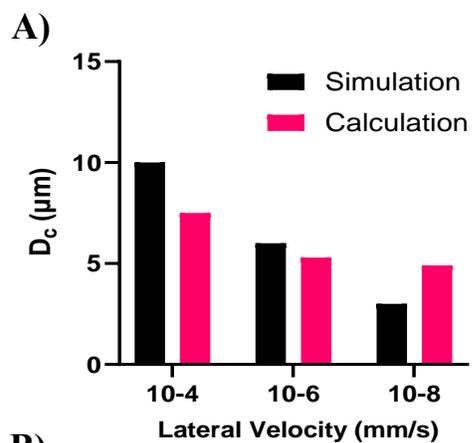

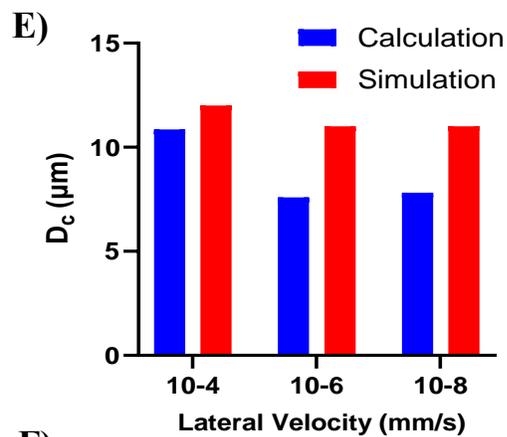

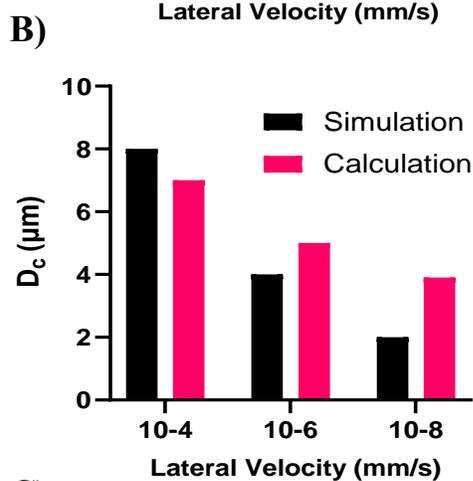

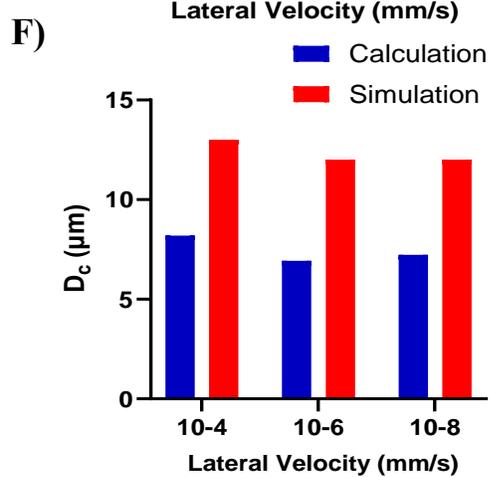

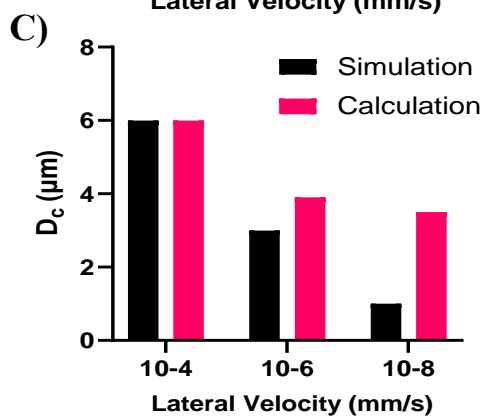

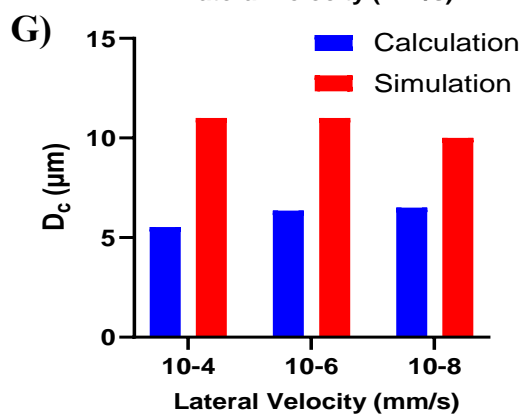

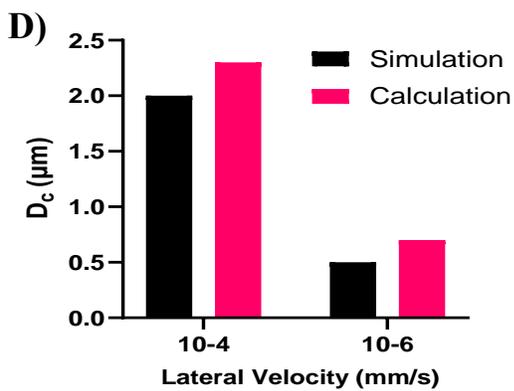

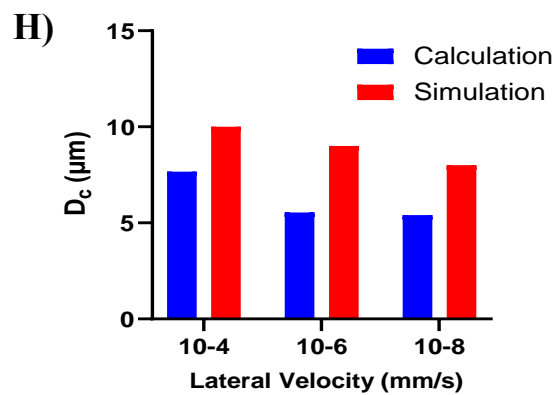

**Figure 10**. **Comparative Analysis of $D_c$ Variations between Python prediction and FEM Simulation**. Comparison of $D_c$ for distinct bypass channel flow rate ratios in the unparallel-slope configuration with bypass channel slopes of **A)** 100-20 μm, **B)** 100-40 μm, **C)** 100-60 μm, and **D)** 100-80 μm. Comparison of $D_c$ for different bypass channel flow rate ratios in the parallel-slope configuration with bypass channel slopes of **E)** 100-20 μm, **F)** 100-40 μm, **G)** 100-60 μm, and **H)** 100-80 μm. It should be mentioned that the velocity in the lower bypass channel is 10 mm/s, whereas the upper one gets velocities of 4, 6, and 8 mm/s respectively. For the cases that 10-8 flow rate ratio is not depicted, the $D_c$ is less than 0.5 $μm$.

# 4. Conclusions

This study delves into the continuous separation of particles using the DLD method and introduces an innovative tunable DLD chip. Employing a numerical model, we examined particle behavior by accounting for all pertinent forces. Additionally, we developed a Python code to predict the $D_c$ based on streamline coordinates prior to determining the exact particles trajectories. Eventually, we offered a comparative analysis between the predicted and exact $D_c$. This novel chip facilitates particle separation in a wide range of $D_c$ values from 0.5 to 14 $μm$, achievable by modifying streamline direction rather than the customary approach of altering pillar array orientation as seen in conventional DLD chips. The chip design showcases a fully horizontal array of pillars, accompanied by two bypass channels positioned at the top and bottom of the DLD chamber. The width of these bypass channels changes linearly from their inlet to their outlet. This chip introduces two design configurations, each distinguished by parallel or unparallel slopes of the bypass channels. Its exceptional attribute lies in its ability to yield a diverse range of $D_c$ values, achieved through the manipulation of two discrete control parameters. The primary control parameter governs the adjustment of flow rates within the bypass channels positioned at the upper and lower sides of the DLD chamber. The second control parameter revolves around the tuning of the slopes of these bypass channels. Both of these factors collectively shape the direction of streamlines carrying particles. By modulating the angle of these streamlines relative to the pillar array, the $D_c$ becomes tunable. Preceding the determination of the $D_c$ for each case, an initial approximation was attained via a Python script employing streamline coordinates. Subsequently, employing FEM modeling for particle trajectories, exact $D_c$ values were derived and subsequently juxtaposed with the initial estimations, revealing minimal discrepancies. Therefore, $D_c$ values can be obtained by having the streamlines obtained from the fluid flow simulation without performing time-consuming FEM modeling to find the particle trajectories. Furthermore, this chip introduces step-by-step particle separation, where differing-sized particles transition from the bumped to zigzag mode at different positions along and across the DLD chamber. This approach exhibits promising outcomes for particle separation across various size ranges, offering high sensitivity across multiple applications. This research holds fundamental significance, addressing challenges posed by geometry and pillar array design that have historically hindered the full realization of DLD chips' potential. By highlighting the notable influence of fluid streamline dynamics on particle separation within a straightforward geometric chip, this study opens avenues for practical implementation.

In the future, our plans include the incorporation of varying gap sizes in the DLD chamber along with the streamline-directed manipulation of $D_c$ within the DLD chamber. This would lead to controllable step-by-step separation across and along the channel. Additionally, an experimental setup to validate this approach is a promising avenue for further exploration.

## 5. Data Availability

The datasets generated and/or analyzed during the current study are not publicly available due to requirement of confidentiality, but are available from the corresponding author on reasonable request.